\documentclass[a4paper,12pt]{article}

\usepackage{natbib}
\usepackage[english]{babel}
\usepackage[latin1]{inputenc}
\usepackage{amsmath, amssymb, theorem}
\usepackage{fullpage}
\usepackage{graphicx}
\usepackage{color}
\usepackage[margin=10pt,font=small]{caption}
\usepackage{bm}
\usepackage{subfig}
\usepackage{setspace}

\usepackage{tikz}
\usetikzlibrary{arrows}
\usepackage{multirow}

\newtheorem{Rem}{\bf Remark}

\newtheorem{Lem}{\bf Lemma}

\def\boxit#1{\vbox{\hrule\hbox{\vrule\kern6pt \vbox{\kern6pt #1\kern6pt}\kern6pt\vrule}\hrule}}

\def\P_25_ICML{{\it Proceedings of the 25th international conference on Machine learning}}

\renewcommand\footnoterule{\kern-3pt \hrule \textwidth 2in \kern 2.6pt}

\numberwithin{equation}{section}
\allowdisplaybreaks

\title{Nonparametric Bayesian Approaches to Non-homogeneous Hidden Markov Models}
\author{Abhra Sarkar\\
Anindya Bhadra\\
Bani K. Mallick}
\date{}

\begin{document}
\maketitle

\begin{abstract}
In this article a flexible Bayesian non-parametric model is proposed for non-homogeneous hidden Markov models.
The model is developed through the amalgamation of the ideas of hidden Markov models and predictor dependent stick-breaking processes. 
Computation is carried out using auxiliary variable representation of the model which enable us to perform exact MCMC sampling from the posterior. 
Furthermore, the model is extended to the situation when the predictors can simultaneously influence the transition dynamics
of the hidden states as well as the emission distribution. 
Estimates of few steps ahead conditional predictive distributions of the response have been used as performance diagnostics for these models. 
The proposed methodology is illustrated through simulation experiments as well as analysis of a real data set concerned with the prediction of rainfall 
induced malaria epidemics.
\vspace{0.3 cm}

\end{abstract}

\section{Introduction}

Hidden Markov models (HMMs), albeit considered to be the simplest forms of Bayesian networks, have been tremendously successful in statistical modeling 
of sequentially generated data with applications in many different areas
like speech recognition \citep{Rabiner:1989, Fox_etal:2008}, 
proteomics \citep{Bae_etal:2005, Lennox_etal:2010}, 
genetics and genomics \citep{Guha_etal:2008, Yau_etal:2011},
economics and finance \citep{Hamilton:1990, Albert_Chib_JBES:1993}. 
\citet{Rabiner:1989, Scott:2002, Yoon:2009} have provided some excellent reviews on HMMs and their applications.

Basic HMM consists of two processes: a hidden process, which evolves according to a first order Markov chain, and an observed process, which is conditionally temporally independent conditioned on the hidden state. 
Let $\bm{y}_{1:T}$ denote potentially multivariate random variables observed sequentially over discrete time points $t=1,2,\dots,T$. 
Let $\bm{z}_{1:T}$ denote the associated sequence of \emph{hidden} states. 
The HMM makes the following set of conditional independence assumptions to model the hidden and the observed processes
\begin{align*}
P(z_t|\bm{z}_{1:t-1})  &= P(z_t|z_{t-1}),	\\
P(\bm{y}_t|\bm{y}_{1:t-1},\bm{z}_{1:t-1}) &= P(\bm{y}_t|z_t).
\end{align*}
Thus it allows the following factorization of the joint distribution of $\big(\bm{y}_{1:T},\bm{z}_{1:T}\big)$
\begin{align*}
P(\bm{y}_{1:T},\bm{z}_{1:T}) = P_0(z_1)P(\bm{y}_1|z_1)\prod_{t=2}^{T}P(z_t|z_{t-1})P(\bm{y}_t|z_t),
\end{align*}
where $P_0$ denotes the distribution of the initial hidden variable $z_1$.  
The observations $\bm{y}_{t}$'s are usually assumed to have been generated according to some parametric probability law $\bm{y}_t|z_t \sim F(\theta_{z_t})$. 
In the context of HMMs, the family of distributions $\{F(\theta_i)\}_i$ is referred to as the family of \emph{emission distributions} 
and the indexing parameters $\{\theta_i\}_i$ are known as the \emph{emission parameters}.
The conditional distributions $P(z_t|z_{t-1})$ governing the evolution of the latent sequence $\{z_{t}\}_{t}$ over time are known as \emph{transition distributions}.

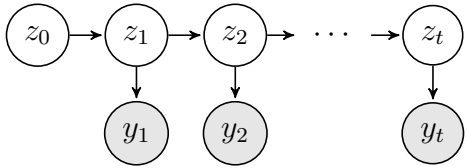
\begin{figure}[h!]
\centering
\begin{tikzpicture}[scale=1.3,->,>=stealth',shorten >=1pt,auto,node distance=2.8cm,semithick]

  \node[style={draw,circle}] (n1) at (0,0) {$z_{0}$};
  \node[style={draw,circle}] (n2) at (1,0) {$z_{1}$};
  \node[style={draw,circle}] (n3) at (2,0) {$z_{2}$};
  \node[style={circle}]      (n4) at (3,0) {$\cdots$};
  \node[style={draw,circle}] (n5) at (4,0) {$z_{t}$};

  \node[style={draw,circle,fill=gray!20}] (n6) at (1,-1) {$y_{1}$};
  \node[style={draw,circle,fill=gray!20}] (n7) at (2,-1) {$y_{2}$};
  \node[style={draw,circle,fill=gray!20}] (n8) at (4,-1) {$y_{t}$};
  
  \path (n1) edge (n2);
  \path (n2) edge (n3);
  \path (n3) edge (n4);
  \path (n4) edge (n5);

  \path (n2) edge (n6);
  \path (n3) edge (n7);
  \path (n5) edge (n8);

\end{tikzpicture}
\caption{Graphical representation of an HMM. Unfilled and shaded nodes signify latent and observable variables respectively.}
\end{figure}

A non-homogeneous hidden Markov model (NHMM) extends this idea by allowing the transition distribution of the hidden states to be dependent on a set of observed 
covariates \citep{Hudges_etal:1999, Shirley_etal:2010}. 
Denoting the observable sequence of potentially multivariate input variables by $\bm{x}_{1:T}$ the conditional independence assumptions for an NHMM can be stated as
\begin{align*}
P(z_t|\bm{z}_{1:(t-1)},\bm{z}_{1:t}) &= P(z_t|z_{t-1},\bm{x}_t),  \\
P(\bm{y}_t|\bm{z}_{1:t}) &= P(\bm{y}_t|z_t),
\end{align*}
with the initial distribution now given by $z_1|\bm{x}_1 \sim P_0(\bm{x}_1)$.
In this situation, the covariates influences the transition dynamics of the latent state variables but 
given $z_t$ the emission distribution does not depend on the value of the covariate. This  model is called NHMM1 in this article.

This model can be further generalized when both the transition and the emission distributions depend on the observed covariates. 
In this frame work, the modeling assumptions  become
\begin{align*}
P(z_t|\bm{z}_{1:(t-1)},\bm{z}_{1:t}) &= P(z_t|z_{t-1},\bm{x}_t),  \\
P(\bm{y}_t|\bm{z}_{1:t},\bm{x}_{1:t} ) &= P(\bm{y}_t|z_t,\bm{x}_t),
\end{align*}
and this extended  model will be referred to as NHMM2.

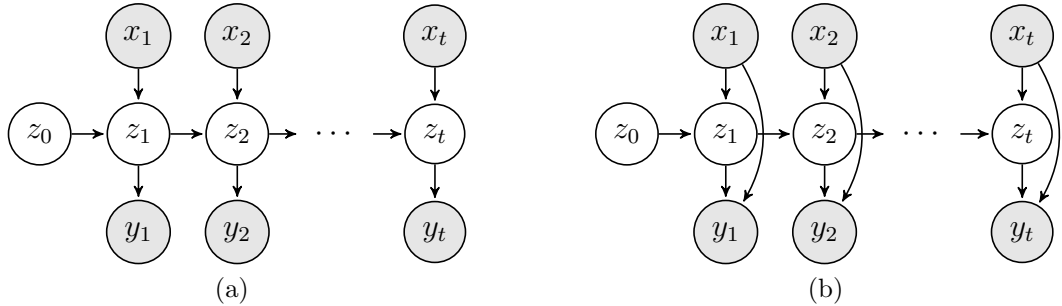
\begin{figure}[h!]
\centering
\subfloat[]{
\begin{tikzpicture}[scale=1.3,->,>=stealth',shorten >=1pt,auto,node distance=2.8cm,semithick]

  \node[style={draw,circle}] (n1) at (0,0) {$z_{0}$};
  \node[style={draw,circle}] (n2) at (1,0) {$z_{1}$};
  \node[style={draw,circle}] (n3) at (2,0) {$z_{2}$};
  \node[style={circle}]      (n4) at (3,0) {$\cdots$};
  \node[style={draw,circle}] (n5) at (4,0) {$z_{t}$};

  \node[style={draw,circle,fill=gray!20}] (n6) at (1,1) {$x_{1}$};
  \node[style={draw,circle,fill=gray!20}] (n7) at (2,1) {$x_{2}$};
  \node[style={draw,circle,fill=gray!20}] (n8) at (4,1) {$x_{t}$};

  \node[style={draw,circle,fill=gray!20}] (n9) at (1,-1) {$y_{1}$};
  \node[style={draw,circle,fill=gray!20}] (n10) at (2,-1) {$y_{2}$};
  \node[style={draw,circle,fill=gray!20}] (n11) at (4,-1) {$y_{t}$};

  \path (n1) edge (n2);
  \path (n2) edge (n3);
  \path (n3) edge (n4);
  \path (n4) edge (n5);

  \path (n6) edge (n2);
  \path (n7) edge (n3);
  \path (n8) edge (n5);

  \path (n2) edge (n9);
  \path (n3) edge (n10);
  \path (n5) edge (n11);

\end{tikzpicture}}
\quad\hspace{1cm}
\subfloat[]{
\begin{tikzpicture}[scale=1.3,->,>=stealth',shorten >=1pt,auto,node distance=2.8cm,semithick]

  \node[style={draw,circle}] (n1) at (0,0) {$z_{0}$};
  \node[style={draw,circle}] (n2) at (1,0) {$z_{1}$};
  \node[style={draw,circle}] (n3) at (2,0) {$z_{2}$};
  \node[style={circle}]      (n4) at (3,0) {$\cdots$};
  \node[style={draw,circle}] (n5) at (4,0) {$z_{t}$};

  \node[style={draw,circle,fill=gray!20}] (n6) at (1,1) {$x_{1}$};
  \node[style={draw,circle,fill=gray!20}] (n7) at (2,1) {$x_{2}$};
  \node[style={draw,circle,fill=gray!20}] (n8) at (4,1) {$x_{t}$};

  \node[style={draw,circle,fill=gray!20}] (n9) at (1,-1) {$y_{1}$};
  \node[style={draw,circle,fill=gray!20}] (n10) at (2,-1) {$y_{2}$};
  \node[style={draw,circle,fill=gray!20}] (n11) at (4,-1) {$y_{t}$};

  \path (n1) edge (n2);
  \path (n2) edge (n3);
  \path (n3) edge (n4);
  \path (n4) edge (n5);

  \path (n6) edge (n2);
  \path (n7) edge (n3);
  \path (n8) edge (n5);

  \path (n2) edge (n9);
  \path (n3) edge (n10);
  \path (n5) edge (n11);

  \path (n6) edge[bend left] (n9);
  \path (n7) edge[bend left] (n10);
  \path (n8) edge[bend left] (n11);


\end{tikzpicture}}
\caption{The two types of NHMM considered in this article. (a) NHMM1: the case when the predictors influence only the transition dynamics,
(b) NHMM2: the more general case when the predictors have direct influence on the emission distribution as well.}
\end{figure}

Direct parallels can be drawn between HMMs and general mixture models. 
Building on ideas of infinite dimensional mixture models, the hierarchical Dirichlet process based HDP-HMM (or iHMM) developed by \citet{Teh_etal:2006} 
paved the way for full Bayesian non-parametric analysis of HMMs. Subsequently, significant  extensions of these models have been proposed by \citet{Van_Gael_etal:2009, Fox_etal:2008}.
Recently in the context of HMM, a Bayesian non-parametric approach was  taken by \citet{Yau_etal:2011} to model genomic copy number variations in array CGH data
in the absence of any covariates. 
They considered an HMM with five states to model a mean function varying slowly over time. 
The residuals were modeled non-parametrically using Dirichlet process mixture.
\cite{Taddy_Kottas:2009} also considered HMM and NHMM models with a finite number of known states and conditionally, given the hidden state, 
non-parametrically modeled the regression relationship between the predictors and the response.

In this article, however, we develop non-parametric models to describe the influence of time and predictors 
on the dynamics of transition of the hidden states.  
A predictor dependent probit stick-breaking process  has been  used to model the transition dynamics of the NHMM as described in the next section. 
Primary emphasis of this article is on the efficient estimation of a few steps ahead predictive densities. 
The article is organized as follows.
Section \ref{Section: Model Specification} introduces the model. 
An exact sampling algorithm for posterior computation is developed in Section \ref{Section: MCMC}.
Section \ref{Section: Prediction} discusses the prediction mechanism.
A simulation study is presented in Section \ref{Section: Simulation Experiments}, 
and an epidemiological application is described in Section \ref{Section: Real Data Analysis}. 
The article concludes with a discussion section.

\section{Model Specification}\label{Section: Model Specification}

An HMM can be treated as a mixture model where the mixing distribution is a Markov chain.
To motivate our model, we, therefore, start with a brief review of Bayesian non-parametric mixtures models. 
In a mixture model, $\bm{y}_{1:n}$ satisfies the conditional independence relation 
\begin{align*}
 P(\bm{y}_{i}|\bm{y}_{-i},\bm{z}_{1:n}) = P(\bm{y}_{i}|z_{i}),
\end{align*}
where $\bm{y}_{-i}$ denotes the set of all observations excluding the $i$-th one and $z_{1:n}$ are associated \emph{hidden} component labels, 
$z_i = k$ implying that the $i$-th observation comes from the $k$-th mixture component.
Component specific parameters, indexed by $z_i$ and additional global parameters, if any, are kept implicit.
In Bayesian non-parametric literature mixture labels $z_i$'s are assigned a prior having countably infinite support $\{1,2,\dots\}$ with 
\emph{random} probability weights associated with its atoms.
Flexibility and richness aside, increasing popularity of these models can be attributed largely to the development of sophisticated computational machinery that has 
made implementation of these techniques routine in various applied problems.


The most celebrated of this type of priors is perhaps the Dirichlet process (DP) prior \citep{Ferguson:1973, Lo:1984, Escobar_West:1995}.
A more general class of infinite dimensional mixing distributions, that includes the DP as a special case,
is the class of stick-breaking priors (SBP) \citep{Sethuraman:1994, Ishwaran_James:2001}
\begin{align*}
P(z_i) &=  \sum_{k=1}^{\infty}\pi_{k}\delta_{k}(z_i),
\end{align*}
where $\pi_k = v_k \prod_{l=1}^{k-1}(1-v_l)$ with $\pi_1 = v_1$ and 
$v_k$'s are independent random variables taking values in the unit interval and $\delta_k$ denotes a point mass at $k$.
Hyper priors on parameters governing the distribution(s) of $v_k$'s allow data to have more influence on the posterior.
Predictor dependent random mixing distributions $\{P(\bm{x}):~ \bm{x}\in \mathcal{X} \}$, where $\mathcal{X}$ denotes the sample space of an associated, 
possibly multivariate predictor variable $\bm{x}$, can be constructed allowing the $v_k$'s to depend on $\bm{x}$. 
Proposals have been plenty, most of them leading to challenging computation \citep{Griffin_Steel:2006, Dunson_Park:2008, Chung_Dunson:2008}.
Probit stick-breaking processes (PSBP) of \citet{Rodriguez_Dunson:2011}, a sub-family of SBP as the nomenclature suggests, is obtained
by setting $v_k = \Phi(\eta_k)$, probit transformation of some underlying random variable $\eta_{k}$ that may depend on an associated predictor, if present.
One such specification that admits easy posterior computation is given by 
\citet{Chung_Dunson:2009} who modeled $v_k(\bm{x}) = \Phi\big(\alpha_k - \sum_{l=1}^{p}\beta_{k,l}|x_{l}-x_{k,l}^{\star}|\big)$,
where $\{x_{k,l}^{\star}\}$ are chosen from a finite set of grid points covering the range of values of $x_l$, 
the $l$-th component of the $p$-variate predictor $\bm{x}$.

In this spirit, we propose a flexible model for NHMM as
\begin{align}
\bm{y}_t|\bm{x}_{t},\bm{\psi},\{\bm{\theta}_j\}_{j=1}^{\infty},z_t &\sim f(\bm{y}_t;\bm{x}_t,\bm{\psi},\bm{\theta}_{z_t}), \label{eq: emission density}  \\
\bm{\theta}_j &\overset{iid}{\sim} p_{0}(\bm{\theta}_j), \label{eq: prior emission parameters 1} \\
\bm{\psi} &\sim p_{0}(\bm{\psi}),  \label{eq: prior emission parameters 2}\\
z_t|z_{t-1}=j, \bm{x}_t &\sim P(j,\bm{x}_t), \label{eq: transition distribution 1}\\
P(j,\bm{x}_t) &= \sum_{k=1}^{\infty}\pi_{k}(j,\bm{x}_t)\delta_{k},  \label{eq: transition distribution 2}\\
\pi_{k}(j,\bm{x}_t) &= \Phi\big(\alpha_{jk}+\beta_k h(\bm{x}_t;\bm{x}_{k}^{\star})\big) \prod_{l=1}^{k-1}\big\{1-\Phi\big(\alpha_{jl}+\beta_l h(\bm{x}_t;\bm{x}_{l}^{\star})\big)\big\}, \label{eq: transition distribution 3} \\
p_{0}(\bm{\alpha},\bm{\beta},\bm{x}^{\star}) &= p_{0}(\bm{\alpha})~p_{0}(\bm{\beta})~p_{0}(\bm{x}^{\star})   \label{eq: prior transition parameters}.
\end{align}
The first equation (\ref{eq: emission density}) describes the emission distribution $f$ which depends on the predictor variables $\bm{x}_{t}$, 
the global parameter $\bm{\psi}$ and component specific parameters $\bm{\theta}_{k}$. 
We assume that $f$ belongs to a parametric class of distributions and assign parametric priors for $\bm{\psi}$ and $\bm{\theta}_{k}$ 
in equations (\ref{eq: prior emission parameters 1}) and (\ref{eq: prior emission parameters 2}). 

At the next hierarchical stage in equation (\ref{eq: transition distribution 1}), we use infinite mixture distribution to model the transition distribution $P$. 
This mixture distribution is given by the stick-breaking process as in equation (\ref{eq: transition distribution 2}). 
More importantly, $P$ depends on the predictors $\bm{x}_t$, and this dependence is induced through the weights $\{\pi_{k}\}_{k}$. 
Accordingly, the $\pi_k$'s are modeled using a probit stick breaking process in equation (\ref{eq: transition distribution 3}) where $\Phi$ is the Gaussian CDF and 
$\alpha_{jk}$, $\beta_{k}$'s are parameters of the probit stick breaking process. 
The function $h(\bm{x}_t;\bm{x}^{\star})$, introduced to model the influence of $\bm{x}_t$ on the state dynamics, 
is specified as 
$h(\bm{x};\bm{x}^{\star}) = -||\bm{x}-\bm{x}_{k}^{\star}||^{2}$, 
where $\{\bm{x}_{k}^{\star}\}_{k}$ are random locations on the predictor space $\mathcal{X}$.
It is assumed here that the components of a multivariate predictor are all standardized to bring them to a common scale.
As $||\bm{x}-\bm{x}_{k}^{\star}|| \rightarrow 0$, $\Phi(\alpha_{jk}+\beta_{k} h(\bm{x},\bm{x}_{k}^{\star})) \rightarrow \Phi(\alpha_{jk})$.
Therefore the maximum probability of transition from state $j$ to state $k$ is attained when $\bm{x}=\bm{x}_{k}^{\star}$.
Restricting the $\beta_{k}$'s to be positive, as $\bm{x}$ goes away from $\bm{x}_{k}^{\star}$, $\Phi(\alpha_{jk}+\beta_{k} h(\bm{x},\bm{x}_{k}^{\star})) \rightarrow 0$, 
i.e. the probability of a transition to the $k$-th state decreases to $0$.
Other parameters remaining fixed, the conditional probability of making a transition from $j$ to $k$ increases with increase in $\alpha_{jk}$. 
Larger values of $\beta_k$ result in faster decay of the probability of a transition to state $k$ as the associated predictor value 
goes away from $\bm{x}_{k}^{\star}$ 
and thereby implies that smaller regions of the predictor space around $\bm{x}_{k}^{\star}$ favor the latent state $k$. 
Finally in equation (\ref{eq: prior transition parameters}), the prior for the parameters of the stick breaking process has been specified as 
$p_{0}(\bm{\alpha},\bm{\beta},\bm{x}^{\star})$.
The above specification does not allow different $\beta_k$ components for different components of a multivariate predictor, leading to a sparse model. 
This restrictive assumption can be relaxed when large number of data points are available.

To specify the initial distribution of $Z_1$, we introduce a special initial state $Z_0$ that is always instantiated at a special value $z_0=0$. 
$P_{0}(\bm{x}_1) = P(0,\bm{x}_1)$ is then specified as $P(0,\bm{x}_1) = \sum_{k=1}^{\infty} \pi_{k}(0,\bm{x}_1)\delta_{k}$,
with $\pi_{k}(0,\bm{x}_1) = \Phi\big(\alpha_{0k}+\beta_k h(\bm{x}_1;\bm{x}_{k}^{\star})\big) \prod_{l=1}^{k-1}\big\{1-\Phi\big(\alpha_{0l}+\beta_l h(\bm{x}_1;\bm{x}_{l}^{\star})\big)\big\}$,
a form that allows inclusion of likelihood contribution from the first output variable 
$\bm{y}_{1}$ in updating $\beta_{1:z_1}$ and $\bm{x}_{1:z_1}^{\star}$ a-posteriori.

The proposed model, when the emission distribution is only implicitly influenced by the associated predictor value $\bm{x}_t$ through $z_t$,
i.e. $f(\bm{y}_t;\bm{x}_t,\bm{\psi},\bm{\theta}_{z_t})= f(\bm{y}_t;\bm{\psi},\bm{\theta}_{z_t})$, 
will be referred to as iNHMM1. 
The more general model, where the predictor directly influences the emission distribution will be referred to as iNHMM2.

\section{Exact MCMC Sampling from the Posterior}\label{Section: MCMC}
This section describes the \emph{exact} MCMC procedure to draw samples from the posterior using auxiliary variables.  The original algorithm for fitting infinite dimensional DPMM by \citet{Escobar_Thesis:1988} and several notable variations of it,
for example, \citet{MacEachern:1994}, \citet{Escobar_West:1995}, \citet{Neal:2000} rely on integrating or `marginalizing' out the random probability measure
and work with the associated Polya urn characterization.
Recent developments have focused on sampling techniques that escape the need of integrating out the random probability measure.
The approximate Gibbs sampler based on truncation by \citet{Ishwaran_James:2001}, the exact retrospective sampler of \citet{Papaspiliopoulos_Roberts:2005} 
and the exact slice sampler of \citet{Walker:2007} and its extension by \citet{Kalli_etal:2011} to a more efficient version are significant contributions
in this direction.
The basic idea in slice samplers is to use random truncation rather than fixed truncation - to limit the space of cluster assignment variables to a random 
but finite size for each MCMC iteration through introduction of auxiliary slice-variables.
By circumventing the necessity to marginalize out the random probability measure, these methods (or their straight-forward extensions)
also allow efficient posterior computation for many different types of stick-breaking processes.

In the context of HMMs, an efficient recursive forward-backward (FB from here onwards) sampler, was originally developed by \citet{Baum_etal:1970} 
for efficient execution of an EM algorithm.
In a Bayesian paradigm direct Gibbs samplers can be implemented for posterior computation in HMMs.
Stochastic versions of the FB sampler, 
lead to an alternative Gibbs sampling strategy that out-performs direct Gibbs sampler in that it results in more rapid mixing and less sample auto-correlation 
\citep{Scott:2002}. 
Unfortunately the FB sampler can not be applied directly to HMMs with infinite state-space.
As in the case of infinite mixture models, approximate sampling techniques based on finite truncation of the state-space and exact 
Gibbs samplers based on marginalization of the random probability measures can be developed for iHMM \citep{Teh_etal:2006, Fox_etal:2008}.
Introducing auxiliary slice-variables, the number of trajectories of latent sequences with positive probabilities for each MCMC iteration can be reduced to a finite size.
Beam sampling, an efficient exact MCMC procedure for drawing samples from the posterior of iHMM, developed by \citet{Van_Gael_etal:2008}, 
builds on this idea and integrates together the FB and the slice sampling techniques.
The procedure can be extended for exact sampling from the posterior of infinite dimensional HMMs with transition distributions constructed through stick-breaking 
processes.

The MCMC procedure to be described here is developed by fusing together modified versions of the FB sampler of \citet{Chib:1996}, 
the slice sampler of \citet{Kalli_etal:2011} and the auxiliary variable sampler of \citet{Chung_Dunson:2009} \citep[see also][]{Albert_Chib:1993}.
More specifically, the latent sequence $\bm{z}_{1:T}$ and the parameters $\{\theta_{j}\}_{j=1}^{\infty}$ specifying the emission distribution 
are updated through a beam sampler and the parameters determining the transition probabilities are updated through an auxiliary variable sampler which is described in the following sections.

\subsection{Introduction of Auxiliary Variables}
We introduce a set of latent variables $\bm{u}_{1:T}$ where the unconditional distribution of each $u_t$ is uniform on the unit interval. 
For any positive sequence $\bm{\xi} = \{\xi_{k}\}_{k=1}^{\infty} \in [0,1]$, we can write 
\begin{align}
 p(u_t,z_t=k|z_{t-1}=j,\pi_k(j,\bm{x}_t),\xi_{k}) = 1(u_t< \xi_{k}) \frac{\pi_k(j,\bm{x}_t)}{\xi_{k}}.\label{eq: conditional prior of u}
\end{align}
The sequence $\{\xi_{k}\}$ is typically a deterministic decreasing sequence, although random sequences are allowed \citep[see][for more details]{Kalli_etal:2011}.
This implies
\begin{align}
 p(u_t|z_t=k,z_{t-1}=j,\pi_k(j,\bm{x}_t),\xi_{k}) &= \frac{p(u_t,z_t=k|z_{t-1}=j,\pi_k(j,\bm{x}_t),\xi_{k})}{p(z_t=k|z_{t-1}=j,\pi_k(j,\bm{x}_t),\xi_{k})} 
	  = \frac{1(u_t< \xi_{k})}{\xi_{k}}, \\
 p(z_t=k|u_t,z_{t-1}=j,\pi_k(j,\bm{x}_t),\xi_{k})  &= \frac{p(u_t,z_t=k|z_{t-1}=j,\pi_k(j,\bm{x}_t),\xi_{k})}{p(u_t|z_{t-1}=j,\pi_k(j,\bm{x}_t),\xi_{k})} \nonumber\\
  &\propto 1(k: u_t < \xi_{k})\frac{\pi_k(j,\bm{x}_t)}{\xi_{k}}.  \label{eq: conditional prior of z}
\end{align}

\noindent Given $\bm{z}_{1:T}$ we also
introduce latent auxiliary variables $\bm{W}^{(z)}=\{W_{z_{t-1},l,t}\}_{l,t=1}^{z_t,T}$ where
\begin{align}
(W_{jk,t}|\alpha_{jk},\beta_k,\bm{x}_t) &\overset{ind}{\sim} N(\alpha_{jk}+\beta_k h(\bm{x}_t;\bm{x}_{k}^{\star}),1),  \\
\{z_t=k|z_{t-1}=j,\bm{x}_t\}  &\text{ iff } \{W_{jk,t} > 0 \text{ and } W_{jl,t} \leq 0 \text{ for } l=1,2,\dots,k-1\}.
\end{align}
In what follows, $\bm{\zeta}$ denotes a generic variable that collects all the parameters that are not explicit.

\subsection{Updating the Latent State Sequence}
The recursive algorithm implemented in this article for updating the latent sequence, is a backward-forward (BF) sampler, a trivial variation of the FB sampler.

\noindent\textbf{$\bullet$ Updating the latent sequence $\bm{u}_{1:T}$: } From equation (\ref{eq: conditional prior of u}), we have,
\begin{align}
p(u_t|\bm{u}^{-t},\bm{z}_{1:T},\bm{y}_{1:T},\bm{x}_{1:T},\bm{\xi},\bm{\zeta}) &= p(u_t|z_t,\bm{\xi}) = \frac{1(0<u_t<\xi_{z_{t}})}{\xi_{z_{t}}}. 
\end{align}

\noindent\textbf{$\bullet$ Updating the latent sequence $\bm{z}_{1:T}$: }\\
Define the backward messages $\beta_t(z_t) = p(\bm{y}_{(t+1):T}|z_t,\bm{x}_{(t+1):T},\bm{u}_{(t+1):T},\bm{\zeta})$, 
with the boundary condition  $\beta_T(z_T) = 1$. 
The following recursion holds -
\begin{align}
\beta_t(z_t) &= p(\bm{y}_{(t+1):T}|z_t,\bm{x}_{(t+1):T},\bm{u}_{(t+1):T},\bm{\zeta}) \nonumber\\
	     &= \sum_{z_{t+1}}p(\bm{y}_{(t+1):T},z_{t+1}|z_t,\bm{x}_{(t+1):T},\bm{u}_{(t+1):T},\bm{\zeta})  \nonumber\\
	     &= \sum_{z_{t+1}}p(\bm{y}_{(t+2):T}|z_{t+1},\bm{x}_{(t+2):T},\bm{u}_{(t+2):T},\bm{\zeta}) ~ p(\bm{y}_{t+1}|z_{t+1},\bm{x}_{t+1},\bm{\zeta}) ~ p(z_{t+1}|z_t,\bm{x}_{t+1},u_{t+1},\bm{\zeta})  \nonumber\\
	     &= \sum_{z_{t+1}}\beta_{t+1}(z_{t+1}) ~ p(z_{t+1}|z_t,\bm{x}_{t+1},u_{t+1},\bm{\zeta}) ~ p(\bm{y}_{t+1}|z_{t+1},\bm{x}_{t+1},\bm{\zeta}).
\end{align}
From equation (\ref{eq: conditional prior of z}), it follows that, given $u_{t+1}$ and the sequence $\{\xi_{k}\}$, 
the set of possible values of $z_{t+1}$, with $p(z_{t+1}|z_t,\bm{x}_{t+1},u_{t+1},\bm{\zeta})>0$, is finite 
and thus the above sum is to be taken only over finitely many values of $z_{t+1}$. 
The joint conditional posterior distribution of the latent states could be factorized as
\begin{align*}
p(\bm{z}_{1:T}|\bm{y}_{1:T},\bm{u}_{1:T},\bm{x}_{1:T},\bm{\zeta}) &= p(z_T|z_{T-1},\bm{y}_{1:T},\bm{u}_{1:T},\bm{x}_{1:T},\bm{\zeta}) \\
            &~~~~~~ \dots ~ p(z_2|z_1,\bm{y}_{1:T},\bm{u}_{1:T},\bm{x}_{1:T},\bm{\zeta}) p(z_1|\bm{y}_{1:T},\bm{u}_{1:T},\bm{x}_{1:T},\bm{\zeta}),
\end{align*}
\begin{align}
\text{where} ~~ p(z_t|z_{t-1},\bm{y}_{1:T},&\bm{x}_{1:T},\bm{u}_{1:T},\bm{\zeta}) \propto p(z_t,\bm{y}_{t:T}|\bm{y}_{1:t-1},z_{t-1},\bm{u}_{1:T},\bm{x}_{1:T},\bm{\zeta})  \nonumber\\
					    &\propto p(\bm{y}_{(t+1):T}|z_t,z_{t-1},\bm{x}_{1:T},\bm{u}_{1:T},\bm{\zeta}) ~ p(y_t|z_t,z_{t-1},\bm{y}_{1:(t-1)},\bm{x}_{1:T},\bm{u}_{1:T},\bm{\zeta}) \nonumber\\
					    & ~~~~~~~~~~~~~~~~~~~~~ \times p(z_t|z_{t-1},\bm{y}_{1:(t-1)},\bm{x}_{1:T},\bm{u}_{1:T},\bm{\zeta})  \nonumber\\
					    &\propto \beta_{t}(z_t) p(\bm{y}_t|z_t,\bm{x}_{t},\bm{\zeta}) p(z_t|z_{t-1},\bm{x}_t,u_t,\bm{\zeta}).
\end{align}
To sample $\bm{z}_{1:T}$ from its full conditional we first pass messages $\beta_t(z_t)$ backwards and then sample forwards.

\subsection{Updating the Parameters of the Emission Distribution}
Conditional posterior distribution of the global parameters $\bm{\psi}$ is given by
\begin{align}
p(\bm{\psi}|\bm{z}_{1:T},\bm{y}_{1:T},\bm{x}_{1:T},\bm{\zeta}) &\propto p_{0}(\bm{\psi}) \prod_{t=1}^{T}f(\bm{y}_{t};\bm{x}_{t},\bm{\psi},\bm{\theta}_{z_t}).
\end{align}
Conditional posterior distribution of cluster specific $\bm{\theta}_k$ is proportional to
\begin{align}
p(\bm{\theta}_{k}|\bm{z}_{1:T},\bm{y}_{1:T},\bm{x}_{1:T},\bm{\zeta}) &\propto p_{0}(\bm{\theta}_k) \prod_{\{t:z_t=k\}}f(\bm{y}_{t};\bm{x}_{t},\bm{\psi},\bm{\theta}_k).
\end{align}
If the set $\{t:z_t=k\}$ is an empty set, i.e. no observation is associated with latent state $k$, then 
$p(\bm{\theta}_{k}|\bm{z}_{1:T},\bm{y}_{1:T},\bm{x}_{1:T},\bm{\zeta})$ is just proportional to its prior.
The conditional posterior can be simplified if the family of distributions $\{p_{0}(\bm{\theta}): \bm{\theta} \in \bm{\Theta}\}$ is conjugate 
for the emission distribution $f(\bm{y}_{t};\bm{x}_{t},\bm{\theta})$.

\subsection{Updating the Parameters of the Transition Distribution}
The parameters of the transition distribution can be updated through Gibbs sampler.

\noindent\textbf{$\bullet$ Updating auxiliary variables: }
For any $W_{jl,t} \in \bm{W}^{(z)}$ we have -
\begin{multline}
(W_{jl,t}|\bm{z}_{1:T},\bm{y}_{1:T},\bm{x}_{1:T},\bm{\zeta}) \sim \big[1(z_{t}=l,z_{t-1}=j) \times 
      N_{+}(w_{jl,t};\alpha_{jl}+\beta_l h(\bm{x}_t;\bm{x}_{l}^{\star}),1) ~ + \\
~~~~~~~~~~ 1(z_{t}>l,z_{t-1}=j) \times N_{-}(w_{jl,t};\alpha_{jl}+\beta_l h(\bm{x}_t;\bm{x}_{l}^{\star}),1) \big],
\end{multline}
where $N_{+}(\mu,\sigma^{2})$ and $N_{-}(\mu,\sigma^{2})$ denote truncated Normal densities with location $\mu$ and scale $\sigma$
truncated below and above zero respectively. 
That is
\begin{align}
p(W_{jl,t}|\bm{z}_{1:T},\bm{y}_{1:T},\bm{x}_{1:T},\bm{\zeta}) \equiv \left\{\begin{array}{c}
       N_{+}(\alpha_{jl}+\beta_l h(\bm{x}_t;\bm{x}_{l}^{\star}),1)  ~~~~~~\text{if }z_{t}=l,z_{t-1}=j   \\
       N_{-}(\alpha_{jl}+\beta_l h(\bm{x}_t;\bm{x}_{l}^{\star}),1)  ~~~~~~\text{if }z_{t}>l,z_{t-1}=j.
       \end{array} \right.
\end{align}

\noindent\textbf{$\bullet$ Updating $\alpha_{jl}$'s: }
This implies -
\begin{multline}
p(\bm{\alpha}|\bm{w},\bm{z}_{1:T};\bm{y}_{1:T};\bm{x}_{1:T};\bm{\zeta}) \propto  \\ 
p_{0}(\bm{\alpha})  \prod_{t=1}^{T}\bigg[\phi(w_{z_{t-1},z_{t},t};\alpha_{z_{t-1},z_{t}}+\beta_{z_{t}} h(\bm{x}_t;\bm{x}_{z_t}^{\star}),1) 
      \bigg\{\prod_{l=1}^{z_t-1}\phi(w_{z_{t-1},l,t};\alpha_{z_{t-1},l}+\beta_l h(\bm{x}_t;\bm{x}_{l}^{\star}),1)\bigg\}\bigg]. \nonumber
\end{multline}

\begin{align}
\begin{split}
\Rightarrow p(\alpha_{jl}|\bm{w},\bm{z}_{1:T},\bm{y}_{1:T},\bm{x}_{1:T},\bm{\zeta}) &\propto p_{0}(\alpha_{jl}) 
      \bigg\{\prod_{\{t:~z_t=l,z_{t-1}=j\}}\phi(w_{jl,t};\alpha_{jl} + \beta_l h(\bm{x}_t;\bm{x}_{l}^{\star}),1)\bigg\}  \\
& \times ~~\bigg\{ \prod_{\{t:~z_t > l,z_{t-1}=j\}} \phi(w_{jl,t};\alpha_{jl} + \beta_l h(\bm{x}_t;\bm{x}_{l}^{\star}),1)\bigg\}       \nonumber \\
&\propto  p_{0}(\alpha_{jl})   \prod_{\{t:~z_t\geq l,z_{t-1}=j\}}\phi(w_{jl,t};\alpha_{jl} + \beta_l h(\bm{x}_t;\bm{x}_{l}^{\star}),1)      \nonumber
\end{split}
\end{align}
If we assume independent Normal prior for $\alpha_{jl}$'s i.e. assume a-priori 
$\alpha_{jl} \overset{ind}{\sim} p_{0}(\alpha_{jl}) \equiv N(\alpha_{jl}; \mu_{\alpha},\sigma_{\alpha}^2)$ then we have
\begin{align}
p(\alpha_{jl}|\bm{w},\bm{z}_{1:T},\bm{y}_{1:T},\bm{x}_{1:T},\bm{\zeta}) \equiv N(\alpha_{jl}~;\mu_{\alpha_{jl} \star},\sigma_{\alpha_{jl} \star}^2),
\end{align}
where $
\mu_{\alpha_{jl} \star} = 
\sigma_{\alpha_{jl} \star}^2 \bigg[ \sum_{\{t:~z_t\geq l,z_{t-1}=j\}} \{w_{jl,t} - \beta_r h(\bm{x}_t;\bm{x}_{r}^{\star})\}+
\frac{\mu_{\alpha}}{\sigma_{\alpha}^2} \bigg],~~
\sigma_{\alpha_{jl} \star}^{-2} = \big(n_{jl}^{\alpha} + \sigma_{\alpha}^{-2}\big),
$ and
$n_{jl}^{\alpha} = \sum_{t=2}^{T} 1(z_t\geq l,z_{t-1}=j)$.

\noindent\textbf{$\bullet$ Updating $\beta_{l}$'s: }
Similarly for the full conditionals of $\beta_l$'s we have -
\begin{align}
\begin{split}
p(\beta_{l}|\bm{w},\bm{z}_{1:T},\bm{y}_{1:T},\bm{x}_{1:T},\bm{\zeta}) &\propto p_{0}(\beta_{l}) 
      \bigg\{\prod_{\{t:~z_t=l\}}\phi(w_{z_{t-1},l,t};\alpha_{z_{t-1},l} + \beta_l h(\bm{x}_t;\bm{x}_{l}^{\star}),1)\bigg\}  \\
& \times ~~\bigg\{ \prod_{\{t:~z_t > l\}} \phi(w_{z_{t-1},l,t};\alpha_{z_{t-1},l} + \beta_l h(\bm{x}_t;\bm{x}_{l}^{\star}),1)\bigg\}       \nonumber \\
&\propto  p_{0}(\beta_{l})   \prod_{\{t:~z_t\geq l\}}\phi(w_{z_{t-1},l,t};\alpha_{z_{t-1},l} + \beta_l h(\bm{x}_t;\bm{x}_{l}^{\star}),1)      \nonumber
\end{split}
\end{align}

\noindent If we assume independent truncated Normal prior for $\beta_{l}$'s i.e. assume a-priori $\beta_l \overset{ind}{\sim} p_{0}(\beta_{l}) \equiv N_{+}(\beta_{l}; \mu_{\beta},\sigma_{\beta}^2)$ then we have
\begin{align}
p(\beta_{l}|\bm{w},\bm{z}_{1:T},\bm{y}_{1:T},\bm{x}_{1:T},\bm{\zeta}) \equiv N_{+}(\beta_l~;\mu_{\beta_l \star},\sigma_{\beta_l \star}^2),
\end{align}
where 
$\mu_{\beta_l \star} = 
      \sigma_{\beta_l \star}^{2} \big\{\sigma_{\beta}^{-2}\mu_{\beta} + \sum_{\{t: z_t \geq l\}} h(\bm{x}_t;\bm{x}_{l}^{\star}) (w_{z_{t-1},l,t} - \alpha_{z_{t-1},l})\big\}$,
$\sigma_{\beta_l \star}^{-2} = \big\{\sigma_{\beta}^{-2} +  \sum_{\{t: z_t \geq l\}}h(\bm{x}_t;\bm{x}_{l}^{\star})^{2} \big\}$.


\noindent\textbf{$\bullet$ Updating $\bm{x}_{l}^{\star}$'s: }
The full conditionals of $\bm{x}_{l}^{\star}$'s are given by -
\begin{multline}
p(\bm{x}^{\star}|\bm{w},\bm{z}_{1:T},\bm{y}_{1:T},\bm{x}_{1:T},\bm{\zeta}) \propto  \\ 
p(\bm{x}^{\star})  \prod_{t=1}^{T}\bigg[\phi(w_{z_{t-1},z_{t},t};\alpha_{z_{t-1},z_{t}} + \beta_{z_t} h(\bm{x}_t;\bm{x}_{z_t}^{\star}),1) 
      \bigg\{\prod_{l=1}^{z_t-1}\phi(w_{z_{t-1},l,t};\alpha_{z_{t-1},l} + \beta_l h(\bm{x}_t;\bm{x}_{l}^{\star}),1)\bigg\}\bigg]. \nonumber
\end{multline}
\begin{align}
\begin{split}
\Rightarrow p(\bm{x}_{l}^{\star}|\bm{w},\bm{z}_{1:T},\bm{y}_{1:T},\bm{x}_{1:T},\bm{\zeta}) &\propto p(\bm{x}_{l}^{\star}) 
      \bigg\{\prod_{\{t:~z_t=l\}}\phi(w_{z_{t-1},l,t};\alpha_{z_{t-1},l} + \beta_l h(\bm{x}_t;\bm{x}_{l}^{\star}),1)\bigg\}  \\
& \times ~~\bigg\{ \prod_{\{t:~z_t > l\}} \phi(w_{z_{t-1},l,t};\alpha_{z_{t-1},l} + \beta_l h(\bm{x}_t;\bm{x}_{l}^{\star}),1)\bigg\}       \nonumber \\
&\propto  p(\bm{x}_{l}^{\star})   \prod_{\{t:~z_t\geq l\}}\phi(w_{z_{t-1},l,t};\alpha_{z_{t-1},l} + \beta_l h(\bm{x}_t;\bm{x}_{l}^{\star}),1).
\end{split}
\end{align}
Assuming $p(\bm{x}_{l}^{\star})$ to be uniform over a discrete set of possible values of $\bm{x}_{l}^{\star}$ on the predictor space $\mathcal{X}$,
the above conditional posterior is a multinomial distribution and therefore can be easily sampled from.

\section{Prediction}\label{Section: Prediction}
Assume that true values of $\bm{x}_{(T+1):(T+n)}$ are known (for example if $\{\bm{x}_t\}$ is a deterministic sequence, or if it affects 
the response series $\{\bm{y}_t\}$ with some time lag $\geq n$) or can be estimated with high precision. 
Collecting the transition and the emission parameters in $\bm{\zeta}$,  
the predictive density for $\bm{y}_{T+n}$ can be written as
\begin{align}
 &f_{T+n}^{pred}(\bm{y}_{T+n}|\bm{y}_{1:T},\bm{x}_{1:(T+n)}) = \int p(\bm{y}_{T+n}|\bm{y}_{1:T},\bm{x}_{1:T},\bm{z}_{1:T},\bm{x}_{(T+1):(T+n)},\bm{\zeta}) ~ dP(\bm{\zeta},\bm{z}_{1:T}|\bm{y}_{1:T},\bm{x}_{1:T})  \nonumber\\
    &= \int \sum_{z_{T+n}} p(\bm{y}_{T+n}|z_{T+n},\bm{x}_{T+n},\bm{\zeta})  \sum_{z_{T+n-1}} p(z_{T+n}|z_{T+n-1},\bm{x}_{T+n},\bm{\zeta})  \sum_{z_{T+n-2}} p(z_{T+n-1}|z_{T+n-2},\bm{x}_{T+n-1},\bm{\zeta})  \nonumber\\
    &~~~~~~~~~~ \cdots ~\sum_{z_{T+1}=1}^{K^{(m)}}p(z_{T+2}|z_{T+1},\bm{x}_{T+2},\bm{\zeta}) ~ p(z_{T+1}|z_{T},\bm{x}_{T+1},\bm{\zeta}) ~ dP(\bm{\zeta},\bm{z}_{1:T}|\bm{y}_{1:T},\bm{x}_{1:T}).
\end{align}
Exact evaluation of the predictive density would be computationally very challenging 
as it involves multiple integral over the parameter space w.r.t. a complex joint posterior density.
Monte Carlo integration techniques, however, can give simple approximation formula.
We use the notation $LHS \hat{=} RHS $ to signify that $LHS$ is being estimated by $RHS$.
Its actual significance being implicitly understood, we will henceforth refer to $f_{T+n}^{pred}(\bm{y}_{T+n}|\bm{y}_{1:T},\bm{x}_{1:(T+n)})$ 
simply as $f_{T+n}^{pred}(\bm{y}_{T+n})$.
A Monte Carlo estimate of $f_{T+n}^{pred}(\bm{y}_{T+n})$ is given by
\begin{align}
   f_{T+n}^{pred}(\bm{y}_{T+n}) ~ &\hat{=} ~ \frac{1}{M}\sum_{m=1}^{M} \sum_{z_{T+n}=1}^{K^{(m)}} p(\bm{y}_{T+n}|z_{T+n},\bm{x}_{T+n},\bm{\zeta}^{(m)}) \sum_{z_{T+n-1}=1}^{K^{(m)}} p(z_{T+n}|z_{T+n-1},\bm{x}_{T+n},\bm{\zeta}^{(m)})   \nonumber\\
	&~~~~~~~~~~  \cdots \sum_{z_{T+1}=1}^{K^{(m)}} p(z_{T+2}|z_{T+1},\bm{x}_{T+2},\bm{\zeta}^{(m)}) p(z_{T+1}|z_T^{(m)},\bm{x}_{T+1},\bm{\zeta}^{(m)})  \nonumber\\
    &= \frac{1}{M}\sum_{m=1}^{M} \hat{f}^{(m)}(\bm{y}_{T+n}|\bm{y}_{1:T},\bm{x}_{1:T},\bm{x}_{(T+1):(T+n)})   = \hat{f}_{T+n}^{pred}(\bm{y}_{T+n}) , ~\text{say},
\end{align}
where $\bm{z}_{1:T}^{(m)}$ and $\bm{\zeta}^{(m)}$ are sampled values of $\bm{z}_{1:T}$ and $\bm{\zeta}$ from $m$-th MCMC iteration, 
and $K^{(m)}= \max \bm{z}_{1:T}^{(m)}$.

\section{Examples}\label{Prior for transition parameters}
The methodology is illustrated through simulation experiments and a real world application.
In Section \ref{Section: Simulation Experiments} predictive performances of the two types of NHMM models are judged using synthetic data sets. 
An epidemiological application is presented in Section \ref{Section: Real Data Analysis}.

For all these examples we specify the hyper-priors as follows. 
Hyper-priors for the parameters of the emission distribution depend on the family the emission distribution comes from and also on the particular application at hand.
Once the hyper-priors for the emission parameters have been specified, to facilitate convergence, 
we recommend that a Dirichlet process mixture model (DPMM) be fitted to the $\bm{y}_t$ values 
with likelihood function the same as the emission distribution but ignoring the time dependence and the influence of the predictors on the latent states.
Latent states $\bm{z}_{1:T}$ could then be initialized at the cluster labels of the DPMM after sufficiently large number of MCMC iterations.
Parameters of the emission distributions can be similarly initialized at the cluster specific parameter estimates from the DPMM.
For univariate $x_t$, the set of possible values of $x_{l}^{\star}$ can be taken to be $\{Q(x,p): p=0.1,0.3,\dots,0.99\}$,
where $Q(x,p)$ denotes the $p$-th percentile of $\bm{x}_{1:T}$.
$x_{k}^{\star}$'s could be instantiated at $\arg \min \sum_{\{t:z_t=k\}}|x_t-x_{l}^{\star}|$.
Based on experience with simulation studies we also recommend that the prior hyper-parameters for $\alpha_{jk}$'s and $\beta_k$'s be set at 
$\mu_{\alpha} = 2, \sigma_{\alpha} = 1, \mu_{\beta} = 2, \sigma_{\beta} = 2/3$. 
The parameters $\alpha_{jk}$'s and $\beta_{k}$'s could all be instantiated at their respective prior means.

\subsection{Simulation Experiments}\label{Section: Simulation Experiments}
\textbf{Simulation Design for iNHMM1: } 
The sequence $x_t$ was generated through an AR(1) process $x_t = 0.95 x_{t-1} + \epsilon_{t}$, where $\epsilon_{t}\sim N(0,1)$. 
$x_t$ values were then standardized.
The state space for the latent variables was $\mathcal{Z} = \{1,2,\dots,5\}$, 
with associated $x^{\star}$ values the $50$-th, $15$-th, $85$-th, $2$-nd and $98$-th percentiles of $\bm{x}_{1:T}$ respectively. 
Transition probabilities were calculated using (\ref{eq: transition distribution 3})( the `$=$' sign now replaced by `$\propto$') 
with parameters $\alpha_{jk} = 2$, if $j=k$ and $\alpha_{jk} = 0.5$, otherwise for all $j,k \in \mathcal{Z}$;  $\beta_{k}=2$ for all $k \in \mathcal{Z}$. 
Emission distribution for $y_t$, given $z_t = k$, was $N(\mu_{y,k},\sigma_{y,k}^2)$ with $\sigma_{y,k}=0.25$ for all $k$ 
and $\mu_{y,k} = 0,-2,2,-4,4$ for $k=1,2,3,4,5$ respectively.
Values of $\sigma_{y,k}$ were all equal to $0.25$, so this parameter could have been treated as a global parameter.
But while fitting the model cluster specific variances were allowed to be different.
Conjugate Normal-Inv-Gamma$(\mu_{0},\sigma_{0}^2/\nu_{0},\gamma_{0},\sigma_{0}^2)$ prior was assigned on $(\mu_{y,k},\sigma_{y,k}^2)$ 
with $\mu_0=0$, $\nu_{0}=0.10$ and $\gamma_{0}$ and $\sigma_{0}^{2}$ were so chosen that the prior mean and sd of $\sigma_{y,k}^{2}$ were $0.20$ and $1$ respectively.
Two different sample sizes, $T=250$ and $T=500$, were considered.
In each case predictor and response values for three additional time points were also simulated. 
The first $T$ data points were used for fitting the model and in each case up to three steps ahead predictive densities were estimated.
A total of $B=100$ data sets were generated using this design. 
$10,000$ MCMC iterations were run in each case and initial $3,000$ iterations were discarded as burn-in.

The mean integrated squared error (MISE) of n-step ahead prediction is defined as 
$MISE = \mathbb{E}\int \{f_{T+n}^{pred}(y)-\hat{f}_{T+n}^{pred}(y)\}^{2}dy$,
which can be estimated by 
\begin{align}
 MISE ~ \hat{=} ~ MISE_{est} = ~ \frac{1}{B}\sum_{b=1}^{B}\sum_{i=1}^{N}\{f_{T+n}^{(b),pred}(y_{i}^{\Delta})-\hat{f}_{T+n}^{(b),pred}(y_{i}^{\Delta})\}^{2}  \Delta_{i},
\end{align}
where $\{y_{i}^{\Delta}\}_{i=0}^{N}$ are a set of grid points on the range of $y$ and 
$\Delta_{i} = (y_{i}^{\Delta} - y_{i-1}^{\Delta})$ for all $i$.
Since a few extreme values can make the estimated MISEs large, $25$-th and $50$-th percentiles are also reported.
The predictive performance is contrasted with that of an infinite dimensional homogeneous HMM, which will be referred to as iHMMP1, 
obtained by replacing all $\beta_k$'s by zeros in iNHMM1. 
The iHMMP1 formulation is based on simple probit stick breaking process and does not model the the influence of the predictor on the state dynamics.

\begin{figure}[!ht]
\begin{center}
 \includegraphics[trim=0cm 7.5cm 0cm 7.0cm, clip=true, scale=0.6]{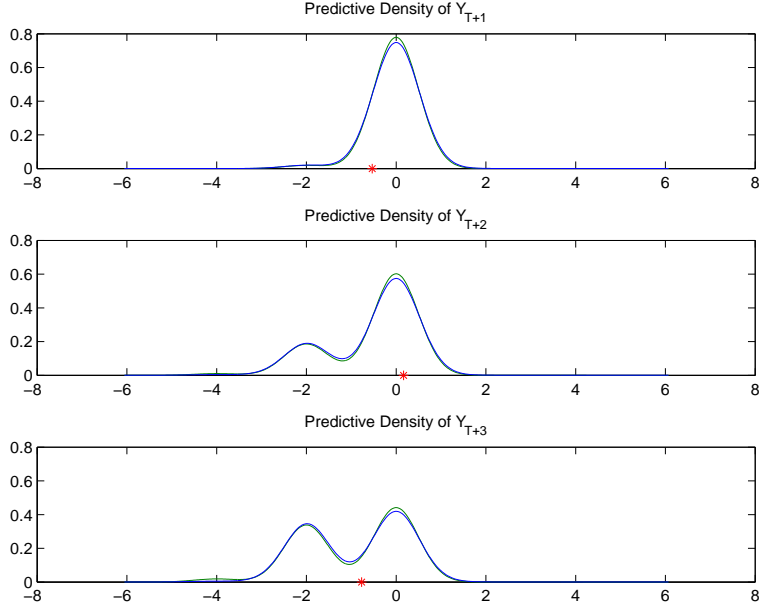}
\end{center}
\caption{Prediction performance of iNHMM1: Three-step ahead true (green) vs estimated (blue) predictive densities for one simulated data set with sample size T$=500$.
	The stars (red) represent true $y$ values.}
\label{fig:PD_iNHMM1_500}
\end{figure}

\begin{table}[ht]\footnotesize
\begin{center}
\begin{tabular}{|c|c|c|c|c|c|c|c|}
\hline
 \multirow{2}{*}{Sample Size} & \multirow{2}{*}{Steps ahead}  & \multicolumn{2}{|c|}{25th Percentile} & \multicolumn{2}{|c|}{50th Percentile} & \multicolumn{2}{|c|}{$MISE_{est}$}\\ \cline{3-8}
			& 	    & iNHMM1 & iHMMP1      & iNHMM1 & iHMMP1      & iNHMM1 & iHMMP1 \\ \hline
\multirow{3}{*}{T = 250}& 1 	    & 0.0019 &  0.0102     & 0.0054 &  0.0638     & 0.0195 & 0.1140      \\
			& 2 	    & 0.0020 &  0.0245     & 0.0067 &  0.0781     & 0.0247 & 0.1696      \\
			& 3 	    & 0.0021 &  0.0309     & 0.0068 &  0.1244     & 0.0296 & 0.2053      \\ \hline
\multirow{3}{*}{T = 500}& 1 	    & 0.0011 &  0.0128     & 0.0039 &  0.0403     & 0.0099 & 0.1068      \\ 
			& 2 	    & 0.0011 &  0.0323     & 0.0050 &  0.0831     & 0.0125 & 0.1655      \\ 
			& 3 	    & 0.0012 &  0.0417     & 0.0050 &  0.1375     & 0.0135 & 0.2037      \\ \hline
\end{tabular}
\caption{Prediction performance of iNHMM1: Summary of one, two and three steps ahead prediction performance for simulation experiments in terms of MISEs}
\label{table:MISEs_iNHMM1}
\end{center}
\end{table}

\noindent\textbf{Simulation Design for iNHMM2: } 
In case of iNHMM2 the emission distribution involves the predictor $x_t$ explicitly.
Number of states, in this case, was fixed at 3 with associated $x^{\star}$ values taken to be the $50$-th, $10$-th and $90$-th percentiles of $\bm{x}_{1:T}$. 
Emission distribution for $y_t$, given $z_t = k$ and $x_t$, was taken to be $N(\eta_{1,k}+\eta_{2,k}x_{t},\sigma_{y}^2)$ with $\sigma_{y}=1$ (global parameter) and 
$\bm{\eta}_{k}^{'} = (\eta_{1,k},\eta_{2,k}) = (1,1), (0,-2), (2,4)$ for $k=1,2,3$ respectively.
In this case $y_t$ values were also standardized before fitting the model.
We assumed $N_{2}(\bm{\eta}_{0},I_{2})$ prior for $\bm{\eta}_{k}$ 
with $\bm{\eta}_{0}$ set at the least square estimate of $\bm{\eta}_{0}$ fitting a simple regression model $y_{t}=\eta_{0}+\eta_{1}x_{t}$.
The prior for $\sigma_{y}^{2}$ was Inv-Gamma$(\gamma_{0},\sigma_{0}^{2})$ with prior mean and sd set at $0.20$ and $1$ respectively.
Again the predictive performance is contrasted with that of a homogeneous HMM, referred to as iHMMP2, 
obtained by replacing all $\beta_k$'s by zeros in iNHMM2.

\begin{figure}[!ht]
\begin{center}
 \includegraphics[trim=0cm 7.5cm 0cm 7.0cm, clip=true, scale=0.6]{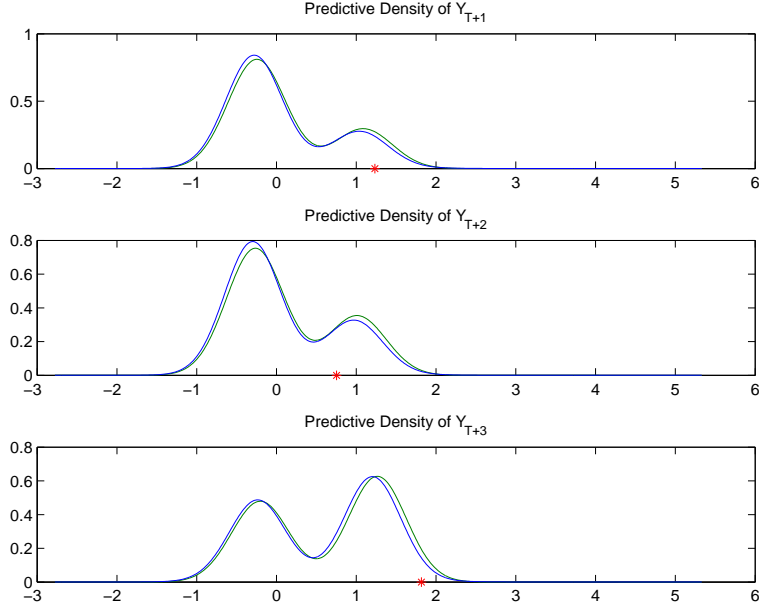}
\end{center}
\caption{Prediction performance of iNHMM2: Three-step ahead true (green) vs estimated (blue) predictive densities for one simulated data set with sample size T$=500$.
	The stars (red) represent true $y$ values.}
\label{fig:PD_iNHMM2_500}
\end{figure}

\begin{table}[ht]\footnotesize
\begin{center}
\begin{tabular}{|c|c|c|c|c|c|c|c|}
\hline
 \multirow{2}{*}{Sample Size} & \multirow{2}{*}{Steps ahead}  & \multicolumn{2}{|c|}{25th Percentile} & \multicolumn{2}{|c|}{50th Percentile} & \multicolumn{2}{|c|}{$MISE_{est}$}\\ \cline{3-8}
			  &         & iNHMM2 & iHMMP2     & iNHMM2 & iHMMP2     & iNHMM2 & iHMMP2 \\ \hline
\multirow{3}{*}{T = 250}  & 1 	    & 0.0025 & 0.0411     & 0.0082 & 0.0920     & 0.0254 & 0.1403     \\
			  & 2 	    & 0.0029 & 0.0678     & 0.0086 & 0.1230     & 0.0247 & 0.1763     \\
			  & 3 	    & 0.0037 & 0.0546     & 0.0110 & 0.1266     & 0.0325 & 0.1906      \\ \hline
\multirow{3}{*}{T = 500}  & 1 	    & 0.0018 & 0.0176     & 0.0048 & 0.0397     & 0.0130 & 0.0959     \\ 
			  & 2 	    & 0.0016 & 0.0273     & 0.0069 & 0.0704     & 0.0131 & 0.1419     \\ 
			  & 3 	    & 0.0021 & 0.0399     & 0.0052 & 0.0998     & 0.0136 & 0.1736      \\ \hline
\end{tabular}
\caption{Prediction performance of iNHMM2: Summary of one, two and three steps ahead prediction performance for simulation experiments in terms of MISEs}
\label{table:MISEs_iNHMM2}
\end{center}
\end{table}

Simulation designs were carefully constructed to ensure diverse shapes of predictive densities.
Figures~\ref{fig:PD_iNHMM1_500} and \ref{fig:PD_iNHMM2_500} represent two such simulation experiments.
The variety of shapes the model can capture should particularly be noted.
Numerical summaries of prediction performance for one, two and three steps ahead prediction are presented in Table~\ref{table:MISEs_iNHMM1} (iNHMM1) 
and Table~\ref{table:MISEs_iNHMM2} (iNHMM2).
From the tables it can be clearly seen that modeling the influence of the predictor on the dynamics of the latent variables, 
produces much improved estimates of the predictive density. MISEs have also been reduced significantly in these situations. 
Although a general increasing trend in MISEs of one, two and three steps ahead predictions may be expected,
since the uncertainty of the predictive distributions also depend on the associated predictor values, this may not always be the case.

\subsection{Application in Prediction of Malaria Epidemics}\label{Section: Real Data Analysis}

\begin{figure}[h]
\begin{center}
\vspace{-5mm}
\includegraphics[height=8cm,width=15cm]{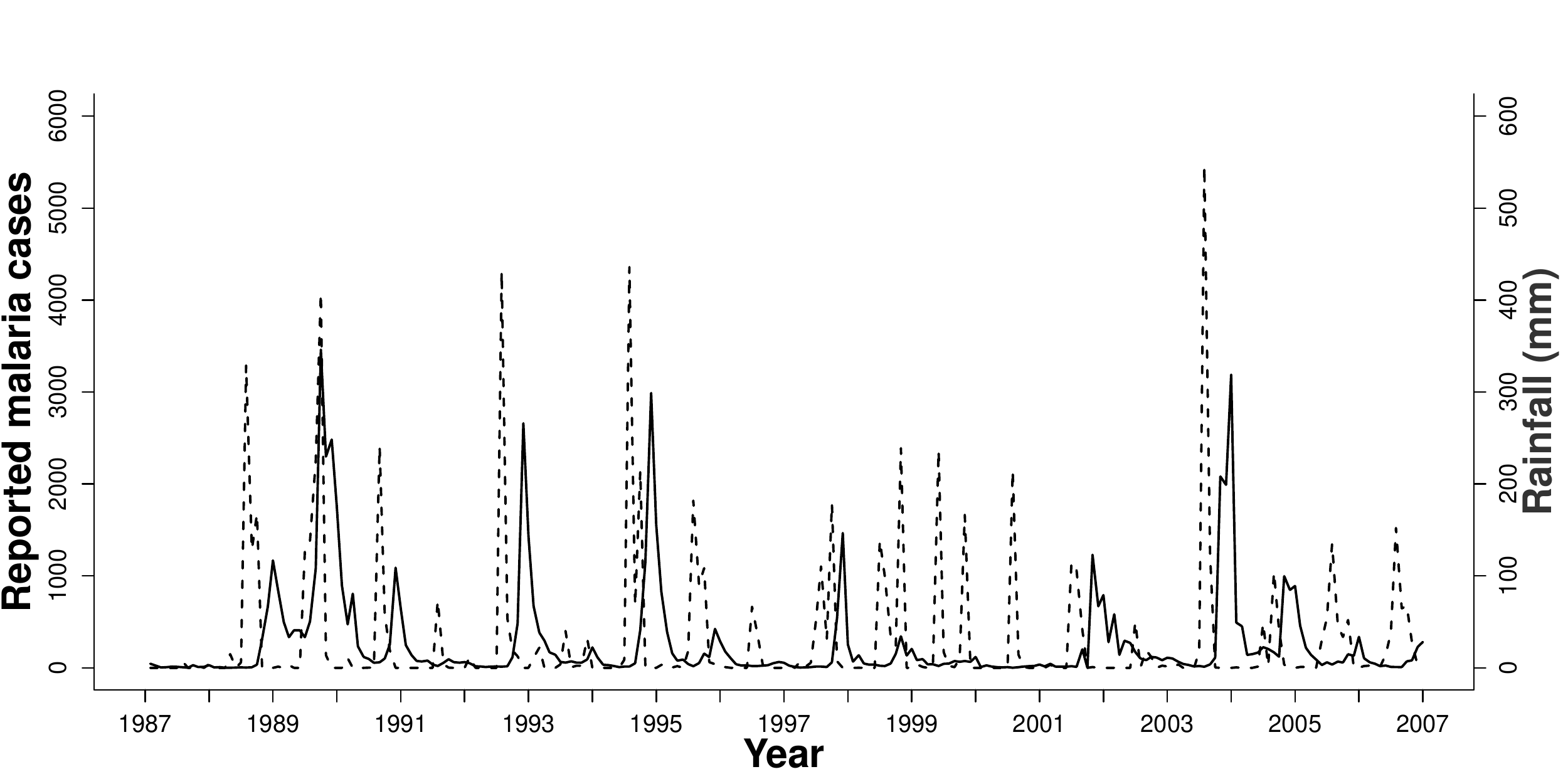}
\vspace{-5mm}
\end{center}
\caption{Monthly reported {\it P.\ falciparum} malaria cases (solid line) and monthly rainfall from a local weather station (broken line) for Kutch, 
	adapted from \cite{Bhadra_etal:2011}.}
\label{fig:cases_rainfall}
\end{figure}

We use the malaria data set used by \cite{Laneri_etal:2010}, \cite{Bhadra_etal:2011} in order to demonstrate the effectiveness of our proposed methodology. 
Figure~\ref{fig:cases_rainfall} displays the monthly confirmed cases of {\it P.~falciparum} and monthly rainfall 
in the district of Kutch, an arid region in the state of Gujarat in Northwest India, between January, 1987 and December, 2006. 
The record of the monthly accumulated malaria cases are maintained by the National Institute of Malaria Research in India, 
and was originally compiled by the office of the District Malaria Officer. 
The monthly accumulated rainfall time series was obtained from a local weather station run by the Indian Meteorology Department.  

\begin{figure}[!ht]
\begin{center}
 \includegraphics[width=16.5cm, height=5cm, trim=2cm 8.4cm 0cm 12.5cm, clip=true]{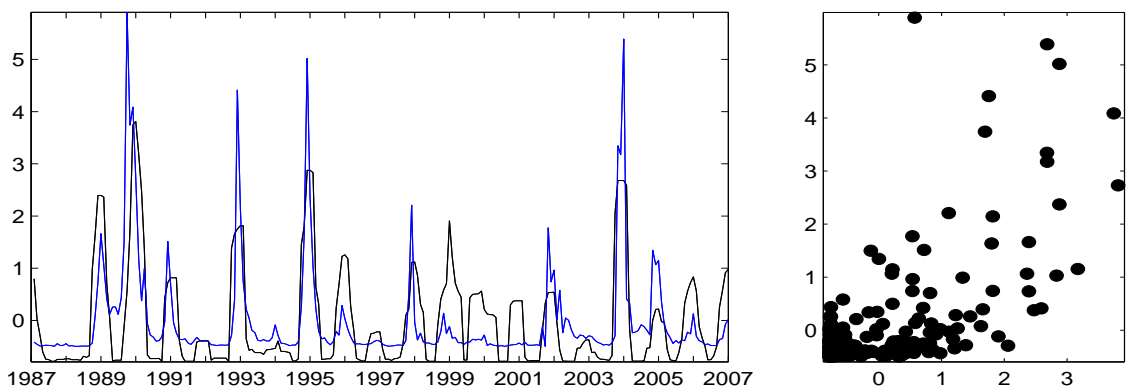}
\end{center}
\caption{Malaria data set: left pane shows the time series of standardized accumulated rainfall (black) and 
	standardized monthly malaria cases (blue) 
	from January, 1987 to December, 2006;
	right pane shows the scatterplot of standardized accumulated rainfall (x-axis) vs standardized monthly malaria cases (y-axis).}
\label{fig:cases_rainfall_2}
\end{figure}

Variability in climate factors can explain a significant share of the variability in regional malaria incidence time series. 
Because the district of Kutch is located in desert region, a region with extreme climate conditions located 
at the edge of the geographical distribution of the disease, 
climate variables such as rainfall are expected to be relevant to disease dynamics \citep{Laneri_etal:2010}.
This malaria time series also shows signs of ``epidemic'' malaria, where the disease peaks in the winter months and typically dies out at the end of the winter. 
This is at a contrast with ``endemic'' malaria where low level infection persists throughout the year.
Visually, an apparent lag relationship between rainfall and reported malaria cases is evident. 
A characteristic of the monsoon climate in this geographic region in India is that the 
rainfall typically peaks during summer monsoon season leading to peaks in malaria several months later during dry winter seasons.
A strong correlation of $0.84$ between total monsoon rainfall (aggregated over June-August) and total winter malaria cases (aggregated over October-December) 
suggests a significant causal relationship \citep{Bhadra_etal:2011}. 
To check the correlation between monthly disease incidence (as opposed to aggregated disease cases over a few months) and the rainfall covariate, 
we used another  window to accumulate rainfall over the past 4 to 6 months from the present month and 
then shifted the accumulated rainfall by a forward lag of 1 or 2 months. 
This resulted in a maximum correlation of 0.72 when we excluded one outlier at September, 1989.

Given these two time series, primary interest lies in the prediction of malaria epidemics for the winter season of a given year given the rainfall covariate until 
the month of September of that year. 
This will enable early preventive measures to be taken, if an epidemic is suspected.
Use of hidden Markov models for modeling disease dynamics and epidemic prediction can be found in the literature.
\cite{Rath_etal:2003}, for example, used an HMM to characterize the non-epidemic and epidemic dynamics in a time series of influenza like disease incidence rates.
See also \citet{Strat_Carrat:1999, Watkins_etal:2009, Conesa_etal:2011}.

We considered all the four models - iNHMM1, iHMMP1, iNHMM2 and iHMMP2, 
with rainfall accumulated over 5 months and shifted forward 2 months as the predictor 
and monthly malaria cases as the response.
Covariate values for the first six months, January to June of 1987, were taken to be the average of that month, averaged over the remaining years.
Henceforth the predictor will simply be referred to as accumulated rainfall.
In our monthly time series of disease cases  spanning over the 20 year period from January, 1987 to December, 2006, we define a particular year to 
be an epidemic year if the accumulated disease cases in that year is greater than the 75-th percentile of aggregated yearly cases, 
where the quantiles are computed based on the data from all of the 20 years (Table 3, \cite{Laneri_etal:2010}).
Predictive performances of the models were evaluated for three different situations - uneventful summer months, winter months of an epidemic year and winter months of a
non-epidemic year.
The models were, thus, fitted to three different subsets of the malaria data set -  
1. first subset consisting of data points from January, 1987 to March, 2006; 
2. a second subset consisting of data points from January, 1987 to September, 2003 (an epidemic year); and 
3. a third subset including data points from January, 1987 to September, 2006.
Recent years 2003 and 2006 were chosen to make the number of data points, used for fitting the models, the maximum available in each case.
In the three cases considered, respectively $T=231, 201$ and $237$ data points were available for model fitting.
Predictor values and responses were standardized and in each case predictive densities were estimated for the following three months.
Thus, in case of the first subset, predictive densities were estimated for the summer months of April, May and June, 
whereas in the latter two cases, predictive densities were estimated for the winter months of October, November and December.
The lag effect of two months implies that $x_{T+1}$ and $x_{T+2}$ are exactly known.
But to calculate $x_{T+3}$, rainfall of the $(T+1)$-th month is required.  
In each case, this was estimated to be the average rainfall of that month calculated from the subset of the data used to fit the models.
Note that the actual predictor is rainfall accumulated over five months and the rainfall of only one component month is required to be estimated.
Also note that the monthly rainfall of the $(T+1)$-th month is actually available from the complete data 
but the models were never allowed to use `future' observations.
For iNHMM1 and iHMMP1 models, given $z_t$, $N(\mu_{y,z_t},\sigma_{y,z_t}^2)$ emission distribution with conjugate but diffuse 
Normal-Inv-Gamma$(2,10,3,1)$ prior for the emission parameters were fitted.
For iNHMM2 and iHMMP2 models, given $z_t$ and $x_t$, $N(\eta_{0,z_t} + \eta_{1,z_t}x_t,\sigma_{y,z_t}^2)$ distribution with non-conjugate diffuse 
$N_2(\bm{\eta}_0,I_2) \times \text{Inv-Gamma}(\gamma_{0},\sigma_{0}^{2})$ priors for $\bm{\eta}_{k} = (\eta_{0,k},\eta_{1,k})^{'}$ and 
$\sigma_{y,k}^{2}$ were fitted.
$\bm{\eta}_{0}$ was set at the least square estimate of $\bm{\eta}_{0}$ fitting a simple regression model $y_{t}=\eta_{0}+\eta_{1}x_{t}$ and 
prior mean and sd of $\sigma_{y,k}^{2}$ were set at $0.5$ and $1$. 
Increasing variability of malaria cases with increase in accumulated rainfall can be seen from Figure \ref{fig:cases_rainfall_2},
prompting us to consider conditionally heteroscedastic emission distributions.
For priors for the parameters of the transition distributions and initialization of the MCMC chain, we refer the reader to 
the beginning of Section~\ref{Prior for transition parameters}.
The scatterplot of accumulated rainfall vs malaria cases does not, however, show any locally varying linear relationship.
Indeed iNHMM1 and iNHMM2 (and similarly iHMMP1 and iHMMP2) models produced very similar results (model fits and predictive densities) in all three situations.
Results for only the more parsimonious iNHMM1 and iHMMP1 models are, therefore, presented.


\begin{figure}[h!]
\begin{center}
 \includegraphics[trim=2.4cm 7.5cm 2.5cm 6.5cm, clip=true, scale=0.4]{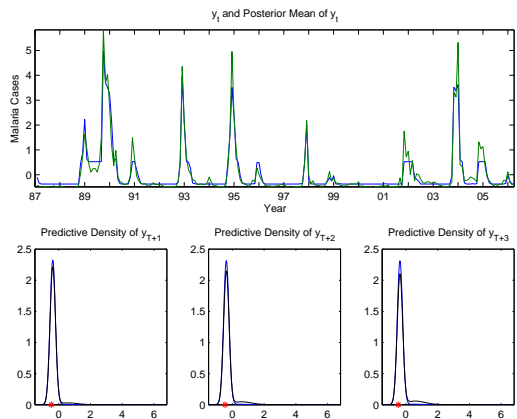}
\end{center}
\caption{Prediction results for summer of 2006: 
      The larger window at the top represent $y_t$ series (green) used to fit the model and the posterior mean sequence (blue) estimated by iNHMM1.
      The three smaller panes at the bottom show the three-step ahead estimated predictive densities for (standardized) $y_t$ series 
      using iNHMM1 (blue) and iHMMP1 (black) for the months of April, May and June of 2006.
      The stars (red) represent true (standardized) $y_t$ values. }
\label{fig:summer_prediction}
\end{figure}

\begin{figure}[h!]
\centering
\subfloat[]{
 \includegraphics[trim=2.4cm 7.5cm 2.5cm 6.5cm, clip=true, scale=0.4]{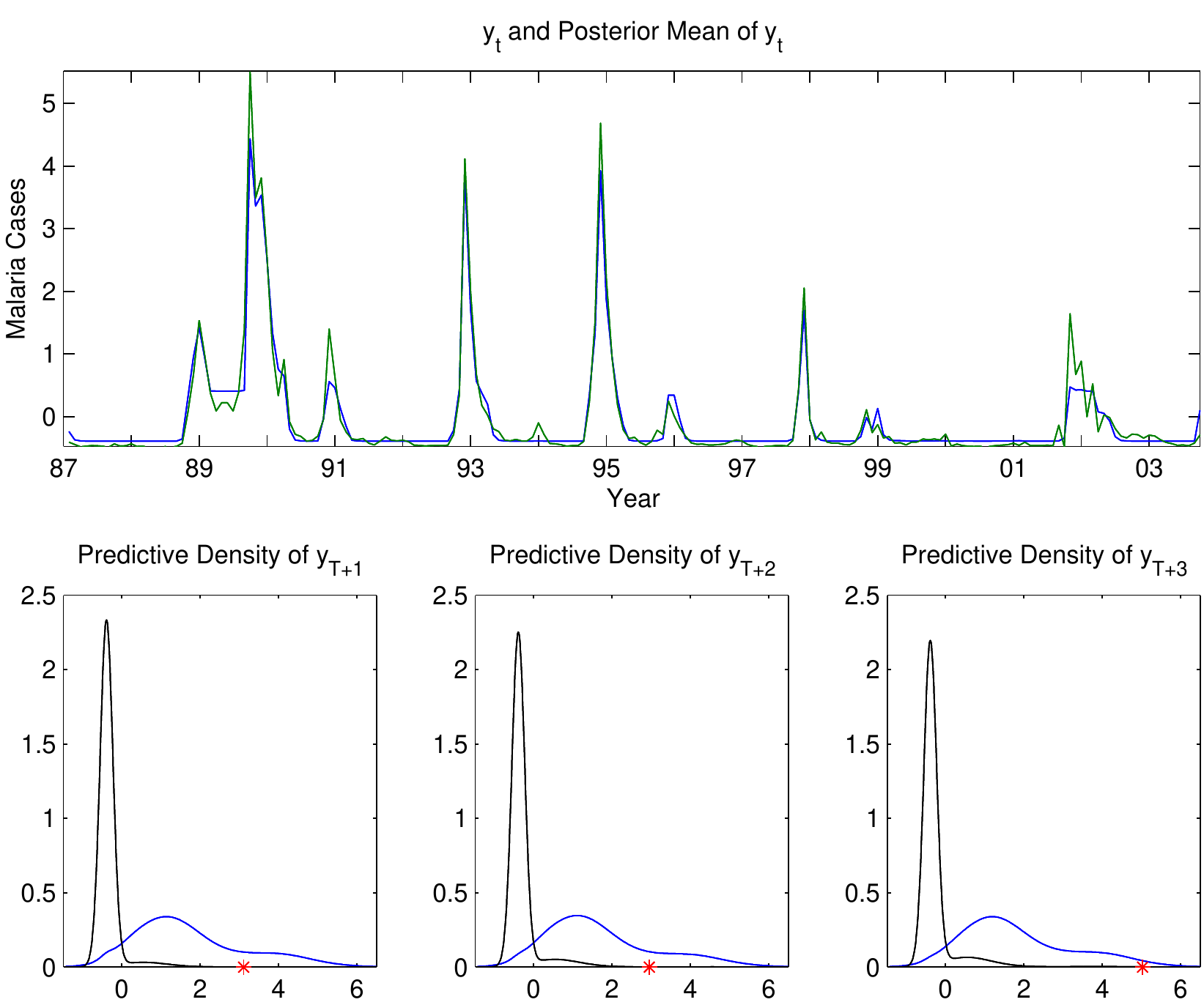}
}
\quad\hspace{1cm}
\subfloat[]{
 \includegraphics[trim=2.4cm 7.5cm 2.5cm 6.5cm, clip=true, scale=0.4]{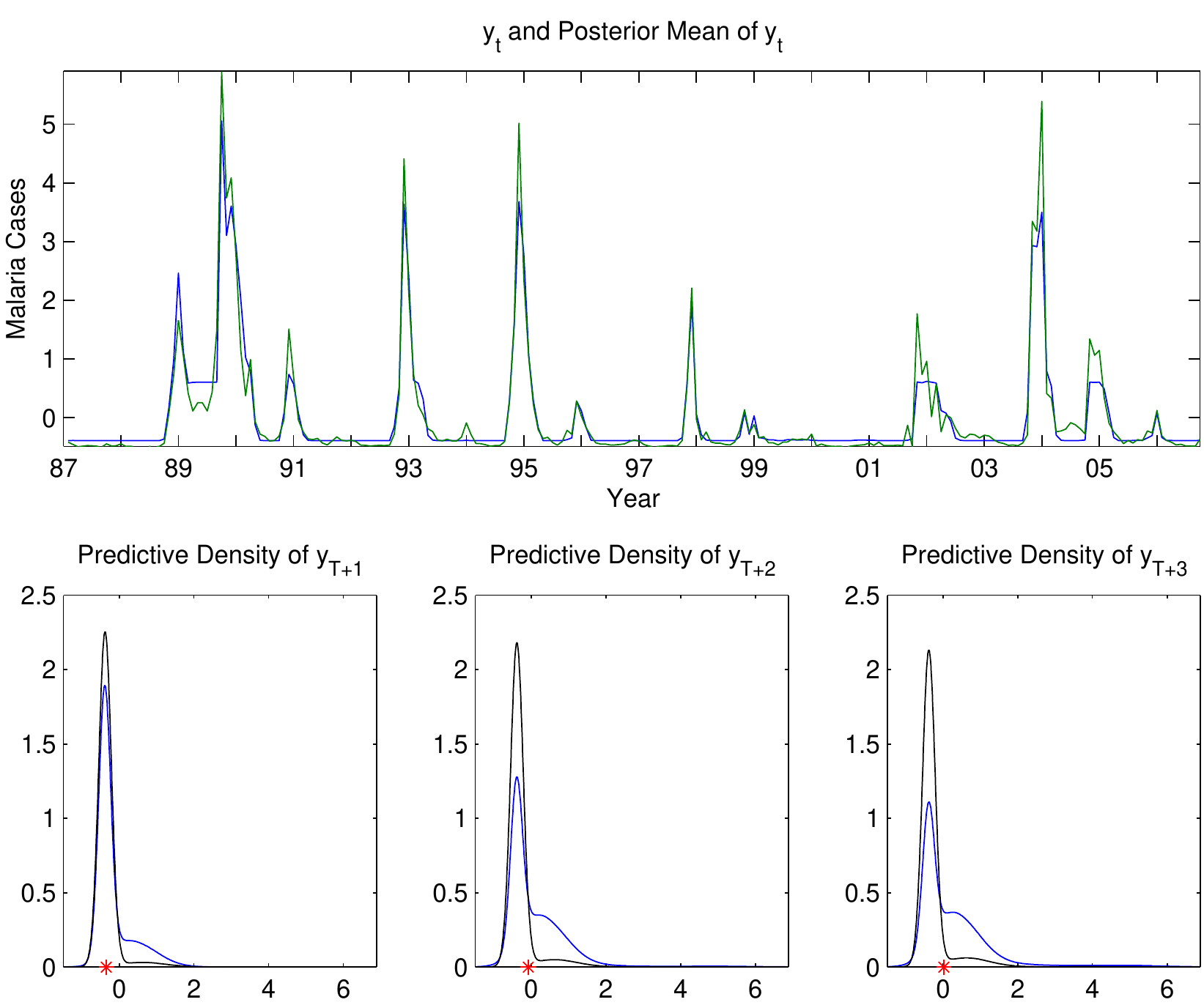}
}
\caption{Prediction results for winters of 2003 and 2006:
      The larger window at the top represent $y_t$ series (green) used to fit the model and the estimated posterior mean (blue) sequence.
      The three smaller panes at the bottom show the three-step ahead estimated predictive densities for (standardized) $y_t$ series 
      using iNHMM1 (blue) and iHMMP1 (black) for the months of October, November and December of (a) 2003 (an epidemic year) and 
      (b) 2006 (a non-epidemic year) respectively.
      The stars (red) represent true (standardized) $y_t$ values. }
\label{fig:winter_prediction}
\end{figure}

As can be seen from the Figure~\ref{fig:summer_prediction} and Figure~\ref{fig:winter_prediction}, for predicting malaria cases for the months of April, May and June of 2006 the models with and without the covariate
produce almost identical results.
However the model without accumulated rainfall as covariate performs poorly in the more important case of estimating the predictive distribution of malaria cases for 
the months of winter (October, November and December).
Because of the presence of only a few sharp peaks in the entire data set, the model without covariate assigns, irrespective of the previous states, 
very small probability of transitions to states that favor large number of monthly malaria cases and large variance.
Conditionally given large values of the predictor, the iNHMM1 model, however, increases the probability of a transition to states favoring 
large number of malaria cases and gives more realistic estimates of the predictive distribution (and associated uncertainty) of malaria cases in winter.

\section{Discussion}
In this article two variations of NHMMs are proposed based on flexible Bayesian non-parametric predictor dependent infinite mixture models. 
Efficient algorithms for exact posterior computation were  developed. 
The proposed methodology is able to produce the full predictive distributions, instead of providing only point prediction estimates.  
Furthermore, this methodology is flexible enough to accommodate multivariate predictors and responses as well as a wide variety of emission distributions including distributions for discrete responses.

The model, introduced in this article, inherits all the strengths and limitations of HMMs and predictor dependent infinite mixture models.
Framework of HMMs, makes the model applicable to situations when the dynamics over time space could be non-linear. 
Use of predictor dependent infinite mixture models, on the other hand, encompasses modeling of scenarios when the change in the shape of the predictive distribution with change in values of the predictor may not follow standard parametric laws. 
Efficient recovery of widely varying predictive densities in simulation experiments and an important epidemiological application illustrate the flexibility and scope of the proposed methodology. 
On the other hand, since the methodology attempts to model dynamical systems with complex dependence relationships between the predictor and the response, 
moderately large number of observations may be required.

The MCMC simulation scheme presented in the paper was exact but does not allow online learning of the dynamical system being modeled.
Inclusion of new data points would necessitate refitting of the models. 
Since the problem of malaria epidemic prediction, described in this paper, required that predictions be made on a monthly basis,
the computational cost of refitting the models was not an issue.
For applications, where online prediction is of importance, sequential Monte Carlo methods can be developed for these models. 
Ongoing and future research projects also include applications of the methodology developed here in the fields of biology and bio-informatics
and an extension to jointly model the transition dynamics and emission distributions within a nonparametric framework.



\section*{Acknowledgments}
This research was supported in part by NSF grant DMS-0914951 and
KUS-C1-016-04 made by King Abdullah University (KAUST).
The authors acknowledge the Texas A\&M University Brazos HPC cluster that contributed to the research reported here.

\bibliographystyle{natbib}
\bibliography{HMM,BayesianNP,MCMC,Epidemics}

\begin{thebibliography}{}

\bibitem[Albert and Chib(1993a)Albert and Chib]{Albert_Chib:1993}
Albert, H. and Chib, S. (1993a).
\newblock {B}ayesian analysis of binary and polychotomous response data.
\newblock {\em \JASA\/}, {\bf 88}, 669--679.

\bibitem[Albert and Chib(1993b)Albert and Chib]{Albert_Chib_JBES:1993}
Albert, J. and Chib, S. (1993b).
\newblock Calculating posterior distributions and modal estimates in {M}arkov
  mixture models.
\newblock {\em \JBES\/}, {\bf 11}, 1--15.

\bibitem[Bae {\em et~al.}(2005)Bae, Mallick, and Elsik]{Bae_etal:2005}
Bae, K., Mallick, B.~K., and Elsik, C.~G. (2005).
\newblock Prediction of protein interdomain linker regions by a hidden {M}arkov
  model.
\newblock {\em \BIOINF\/}, {\bf 21}(10).

\bibitem[Baum {\em et~al.}(1973)Baum, Petrie, Soules, and
  Weiss]{Baum_etal:1970}
Baum, L., Petrie, T., Soules, G., and Weiss, N. (1973).
\newblock A maximizing technique occurring in the statistical analysis of
  probabilistic functions of {M}arkov chains.
\newblock {\em \ANNALSMS\/}, {\bf 41}, 164--171.

\bibitem[Bhadra {\em et~al.}(2011)Bhadra, Ionides, Laneri, Bouma, Dhiman, and
  Pascual]{Bhadra_etal:2011}
Bhadra, A., Ionides, E.~L., Laneri, K., Bouma, M., Dhiman, R.~C., and Pascual,
  M. (2011).
\newblock Forcing versus feedback: Epidemic malaria and monsoon rains in
  northwest {I}ndia.
\newblock {\em \JASA\/}, {\bf 106}(494), 440--451.

\bibitem[Chib(1996)Chib]{Chib:1996}
Chib, S. (1996).
\newblock Calculating posterior distributions and modal estimates in {M}arkov
  mixture models.
\newblock {\em \JECM\/}, {\bf 75}, 79--97.

\bibitem[Chung and Dunson(2008)Chung and Dunson]{Chung_Dunson:2008}
Chung, Y. and Dunson, D. (2008).
\newblock The local {D}irichlet processes.
\newblock {\em \ANNALSISM\/}, {\bf 63}(1), 59--80.

\bibitem[Chung and Dunson(2009)Chung and Dunson]{Chung_Dunson:2009}
Chung, Y. and Dunson, D.~B. (2009).
\newblock Nonparametric {B}ayes conditional distribution modeling with variable
  selection.
\newblock {\em \JASA\/}, {\bf 104}(488), 1646--1660.

\bibitem[Conesa {\em et~al.}(2011)Conesa, Martinez-Beneito, Amoros, and
  Lopez-Quilez]{Conesa_etal:2011}
Conesa, D., Martinez-Beneito, M.~A., Amoros, R., and Lopez-Quilez, A. (2011).
\newblock Bayesian hierarchical {P}oisson models with a hidden {M}arkov
  structure for the detection of influenza epidemic outbreaks.
\newblock {\em \SMMR\/}.
\newblock doi: 10.1177/0962280211414853.

\bibitem[Dunson and Park(2008)Dunson and Park]{Dunson_Park:2008}
Dunson, D. and Park, J. (2008).
\newblock Kernel stick-breaking processes.
\newblock {\em \BIOK\/}, {\bf 95}(1), 859--874.

\bibitem[Escobar(1988)Escobar]{Escobar_Thesis:1988}
Escobar, M.~D. (1988).
\newblock Estimating the means of several {N}ormal populations by
  non-parametric estimation of the distribution of the means.
\newblock Unpublished Ph.D. thesis, Department of Statistics, Yale University.

\bibitem[Escobar and West(1995)Escobar and West]{Escobar_West:1995}
Escobar, M.~D. and West, M. (1995).
\newblock Bayesian density estimation and inference using mixtures.
\newblock {\em \JASA\/}, {\bf 90}(430), 577--588.

\bibitem[Ferguson(1973)Ferguson]{Ferguson:1973}
Ferguson, T.~F. (1973).
\newblock A {B}ayesian analysis of some nonparametric problems.
\newblock {\em \ANNALS\/}, {\bf 1}(2), 209--230.

\bibitem[Fox {\em et~al.}(2008)Fox, Sudderth, Jordan, and
  Willsky]{Fox_etal:2008}
Fox, E.~B., Sudderth, E.~B., Jordan, M.~I., and Willsky, A.~S. (2008).
\newblock An {HDP-HMM} for systems with state persistence.
\newblock {\em \P_25_ICML\/}.

\bibitem[Griffin and Steel(2006)Griffin and Steel]{Griffin_Steel:2006}
Griffin, J.~E. and Steel, M. F.~J. (2006).
\newblock Order-based dependent {D}irichlet processes.
\newblock {\em \JASA\/}, {\bf 101}(473), 179--194.

\bibitem[Guha {\em et~al.}(2008)Guha, Li, and NewBerg]{Guha_etal:2008}
Guha, S., Li, Y., and NewBerg, D. (2008).
\newblock Bayesian hidden {M}arkov modeling of array {CGH} data.
\newblock {\em \JASA\/}, {\bf 103}(482), 485--497.

\bibitem[Hamilton(1990)Hamilton]{Hamilton:1990}
Hamilton, J.~E. (1990).
\newblock Analysis of time series subject to changes in regime.
\newblock {\em \JECM\/}, {\bf 45}, 39--70.

\bibitem[Hudges {\em et~al.}(1999)Hudges, Guttorp, and
  Charles]{Hudges_etal:1999}
Hudges, J.~P., Guttorp, P., and Charles, S.~P. (1999).
\newblock A non-homogeneous hidden {M}arkov model for precipitation occurrence.
\newblock {\em \JRSSC\/}, {\bf 48}(1), 15--30.

\bibitem[Ishwaran and James(2001)Ishwaran and James]{Ishwaran_James:2001}
Ishwaran, H. and James, L.~F. (2001).
\newblock Gibbs sampling methods for stick-breaking priors.
\newblock {\em \JASA\/}, {\bf 96}(453), 161--173.

\bibitem[Kalli {\em et~al.}(2011)Kalli, Griffin, and Walker]{Kalli_etal:2011}
Kalli, M., Griffin, J.~E., and Walker, S.~G. (2011).
\newblock Slice sampling mixture models.
\newblock {\em \SaC\/}, {\bf 21}(1), 93--105.

\bibitem[Laneri {\em et~al.}(2010)Laneri, Bhadra, Ionides, Bouma, Yadav,
  Dhiman, and Pascual]{Laneri_etal:2010}
Laneri, K., Bhadra, A., Ionides, E.~L., Bouma, M., Yadav, R., Dhiman, R., and
  Pascual, M. (2010).
\newblock Forcing versus feedback: Epidemic malaria and monsoon rains in
  northwest {I}ndia.
\newblock {\em \PLoSCB\/}, {\bf 6}(9).
\newblock e1000898.

\bibitem[Lennox {\em et~al.}(2010)Lennox, Dahl, Day, and
  Tsai]{Lennox_etal:2010}
Lennox, K.~P., Dahl, D.~B., Day, R., and Tsai, W. (2010).
\newblock A {D}irichlet process mixture of hidden {M}arkov models for protein
  structure prediction.
\newblock {\em \ANNALSAS\/}, {\bf 4}(2), 916--942.

\bibitem[Lo(1984)Lo]{Lo:1984}
Lo, A.~Y. (1984).
\newblock On a class of {B}ayesian nonparametric estimates. {I}: Density
  estimates.
\newblock {\em \ANNALS\/}, {\bf 12}(1), 351--357.

\bibitem[MacEachern(1994)MacEachern]{MacEachern:1994}
MacEachern, S. (1994).
\newblock Estimating {N}ormal means with a conjugate style {D}irichlet process
  prior.
\newblock {\em \COMMSS\/}, {\bf 23}(3), 727--741.

\bibitem[Neal(2000)Neal]{Neal:2000}
Neal, R.~M. (2000).
\newblock Markov chain sampling methods for {D}irichlet process mixture models.
\newblock {\em \JCGS\/}, {\bf 9}(2), 249--265.

\bibitem[Papaspiliopoulos and Roberts(2005)Papaspiliopoulos and
  Roberts]{Papaspiliopoulos_Roberts:2005}
Papaspiliopoulos, O. and Roberts, O. (2005).
\newblock Retrospective {M}arkov chain monte-carlo methods for {D}irichlet
  process hierchical mixture models.
\newblock {\em \BIOK\/}, {\bf 95}, 169--186.

\bibitem[Rabiner(1989)Rabiner]{Rabiner:1989}
Rabiner, L. (1989).
\newblock A tutorial on hidden {M}arkov models and selected applications in
  speech recognition.
\newblock {\em \IEEE\/}, {\bf 77}, 257--286.

\bibitem[Rath {\em et~al.}(2003)Rath, Carreras, and Sebastiani]{Rath_etal:2003}
Rath, T.~M., Carreras, M., and Sebastiani, P. (2003).
\newblock Automated detection of influenza epidemics with hidden {M}arkov
  models.
\newblock In {\em Proceedings of IDA'03\/}, pages 521--532.

\bibitem[Rodriguez and Dunson(2011)Rodriguez and Dunson]{Rodriguez_Dunson:2011}
Rodriguez, A. and Dunson, D. (2011).
\newblock Nonparametric {B}ayesian models through probit stick-breaking
  processes.
\newblock {\em \BA\/}, {\bf 6}(1), 145--178.

\bibitem[Scott(2002)Scott]{Scott:2002}
Scott, S.~L. (2002).
\newblock Bayesian methods for hidden {M}arkov models recursive computing in
  the 21st century.
\newblock {\em \JASA\/}, {\bf 97}(457), 337--351.

\bibitem[Sethuraman(1994)Sethuraman]{Sethuraman:1994}
Sethuraman, J. (1994).
\newblock A constructive definition of {D}irichlet priors.
\newblock {\em \SSNC\/}, {\bf 4}, 639--650.

\bibitem[Shirley {\em et~al.}(2010)Shirley, Small, Lynch, Maisto, and
  Oslin]{Shirley_etal:2010}
Shirley, K.~E., Small, D.~S., Lynch, K.~G., Maisto, S.~A., and Oslin, D.~W.
  (2010).
\newblock Hidden {M}arkov models for alcoholism treatment trial data.
\newblock {\em \ANNALSAS\/}, {\bf 4}(1), 366--395.

\bibitem[Strat and Carrat(1999)Strat and Carrat]{Strat_Carrat:1999}
Strat, Y.~L. and Carrat, F. (1999).
\newblock Monitoring epidemiologic surveillance data using hidden {M}arkov
  model.
\newblock {\em \STATMED\/}, {\bf 18}(24).

\bibitem[Taddy and Kottas(2009)Taddy and Kottas]{Taddy_Kottas:2009}
Taddy, M.~A. and Kottas, A. (2009).
\newblock Markov switching {D}irichlet process regression.
\newblock {\em \BA\/}, {\bf 4}(4), 793--816.

\bibitem[Teh {\em et~al.}(2006)Teh, Jordan, J., and M.]{Teh_etal:2006}
Teh, Y.~W., Jordan, M.~I., J., B.~M., and M., B.~D. (2006).
\newblock Hierarchical {D}irichlet processes.
\newblock {\em \JASA\/}, {\bf 101}(476), 1566--1581.

\bibitem[Van~Gael {\em et~al.}(2008)Van~Gael, Saatci, Teh, and
  Ghahramani]{Van_Gael_etal:2008}
Van~Gael, J., Saatci, Y., Teh, Y.~W., and Ghahramani, Z. (2008).
\newblock Beam sampling for the infinite hidden {M}arkov model.
\newblock {\em \P_25_ICML\/}.

\bibitem[Van~Gael {\em et~al.}(2009)Van~Gael, Teh, and
  Ghahramani]{Van_Gael_etal:2009}
Van~Gael, J., Teh, Y.~W., and Ghahramani, Z. (2009).
\newblock The infinite factorial hidden {M}arkov model.
\newblock {\em \ANIPS\/}, {\bf 21}, 1697--1704.

\bibitem[Walker(2007)Walker]{Walker:2007}
Walker, S.~G. (2007).
\newblock Sampling the {D}irichlet mixture model with slices.
\newblock {\em \COMMSSC\/}, {\bf 36}, 45--54.

\bibitem[Watkins {\em et~al.}(2009)Watkins, Eagleson, Veenendaal, and
  Wright]{Watkins_etal:2009}
Watkins, R.~E., Eagleson, S., Veenendaal, B., and Wright, G. (2009).
\newblock Disease surveillance using a hidden {M}arkov model.
\newblock {\em \BMCMIDM\/}, {\bf 9}(39).

\bibitem[Yau {\em et~al.}(2011)Yau, Papaspiliopoulos, Roberts, and
  Holmes]{Yau_etal:2011}
Yau, C., Papaspiliopoulos, O., Roberts, G.~O., and Holmes, C. (2011).
\newblock Byesian non-parametric hidden {M}arkov models with applications in
  genomics.
\newblock {\em \JRSSB\/}, {\bf 73}(1), 37--57.

\bibitem[Yoon(2009)Yoon]{Yoon:2009}
Yoon, B.~J. (2009).
\newblock Hidden {M}arkov models and their applications in biological sequence
  analysis.
\newblock {\em \CG\/}, {\bf 10}(6), 402--415.

\end{thebibliography}

\end{document}